\documentclass{article}
\usepackage{graphicx}
\usepackage{amssymb}
\newcommand{\be}{\begin{equation}}
\newcommand{\ee}{\end{equation}}
\newcommand{\bea}{\begin{eqnarray}}
\newcommand{\eea}{\end{eqnarray}}
\newcommand{\nn}{\nonumber \\}
\newcommand{\ex}[1]{{\rm e}^{#1}}
\newcommand{\half}{1/2}
\newcommand{\zero}{\equiv 0 \: {\rm mod} \: 2}
\newcommand{\dg}{^\dagger}
\newcommand{\de}{{\rm d}}
\newcommand{\ie}{{\rm i}}
\newcommand{\hte}{{\it HTE}\,}
\newcommand{\lte}{{\it LTE}\,}
\newcommand{\vau}{{\upsilon}}
\newcommand{\nicht}[1]{ }
\newcommand{\Zm}{\{0,1\}}
\begin{document}

\title{Duality in generalized Ising models}

\author{Franz J. Wegner\\
Institut f\"ur Theoretische Physik,
Ruprecht-Karls-Universit\"at Heidelberg}

\maketitle

\begin{center}
Contribution to the summer school\\
'Topological aspects of condensed matter physics'\\
Les Houches August 2014
\end{center}

{\bf Abstract} This paper rests to a large extend on a paper I wrote
some time ago on {\it Duality in generalized Ising models and phase 
transitions without local order parameter}. It deals with Ising models
with interactions containing products of more than two spins.
In contrast to this old paper I will first give examples before I come
to the general statements.

Of particular interest is a gauge-invariant Ising model in four dimensions.
It has important
properties in common with models for quantum chromodynamics as
developed by Ken Wilson. One phase yields an area law for the Wilson-loop
yielding an interaction increasing proportional to the distance and
thus corresponding to quark-confinement. The other phase yields a perimeter law
allowing for a quark-gluon plasma.

\section{Introduction}

In this contribution I consider a number of Ising models, which arose out of the
question, whether there is duality for Ising models in dimensions larger than
two.
Indeed the idea of duality can be used to construct a whole class of such
systems,
which however, differ from conventional Ising models in some properties.
First these models contain interactions with products of more than two Ising
spins.
Secondly they have
no longer local order parameters, but they can still have two phases.
For a number of these systems the order appears in the expectation value of the
product of the spins along a loop, called Wilson-loop. It shows in the limit of
large loops an area law at high temperatures and a perimeter law at low
temperatures.

Such models, where the Ising spins are replaced by elements of groups, typically
by the groups U(1), SU(2) and SU(3), have become important as lattice gauge
models in high-energy physics for the description of quarks and gluons.

In Section \ref{KraWan} I review the Kramers-Wannier duality for two-dimensional
Ising models. In Section \ref{threedim} I introduce the model dual to the
conventional
three-dimensional Ising model.
Section \ref{gen} inroduces the general concept of Ising models
and duality.
In Section \ref{Mdn} this is applied to general lattices and in
Section \ref{Mdncub} to models on hypercubic lattices. The correlation functions
are considered in Section \ref{correl}.
The basic idea of lattice gauge theory is
given in Section \ref{latt} and a useful lattice for the discretization of
Maxwell's equations is
mentioned in Section \ref{electro}.

\section{Kramers-Wannier Duality}
\label{KraWan}

Kramers and Wannier\cite{KraWan41,Wannier45} predicted in 1941 the exact
critical
temperature of the two-dimensional Ising model on a square lattice.
They did this by comparing the high- and the low-temperature expansion
for the partition function of this model.
Consider a square lattice with $N_s=N_1\times N_2$ lattice points and periodic
boundary conditions. There is an Ising spin $S_{i,j}=\pm 1$ at each lattice
site. The Hamiltonian reads
\be
H=-J\sum_{i=1}^{N_1}\sum_{j=1}^{N_2} (S_{i,j}S_{i,j+1} + S_{i,j}S_{i+1,j}).
\ee
\smallskip

\begin{figure}
\begin{center}
\includegraphics[scale=0.6]{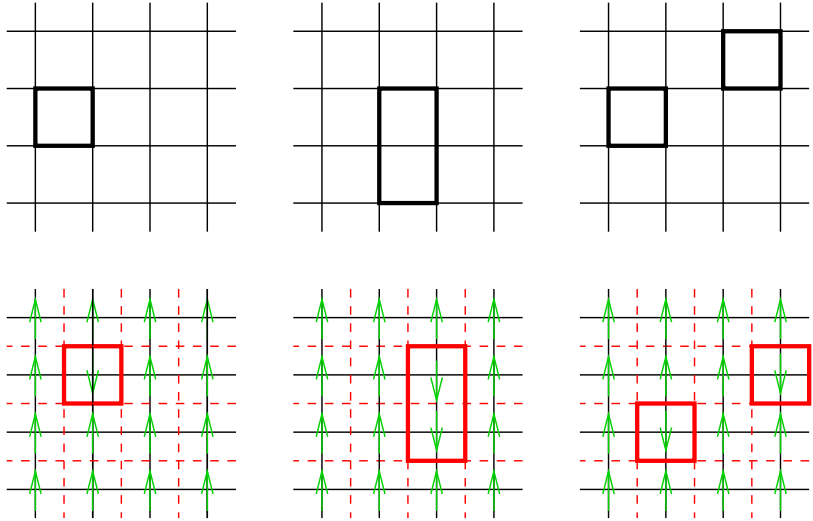}
\end{center}
\caption{Examples for closed loops in the \hte\ and Bloch walls
in the \lte\ on the dual lattice}
\label{Ising_square}
\end{figure}

\noindent
{\bf High temperature expansion} (\hte) We may rewrite the Boltzmann factor
\bea
\ex{-\beta H} &=& \prod_{i,j} (\cosh K+\sinh K S_{i,j}S_{i,j+1})
(\cosh K+\sinh K S_{i,j}S_{i+1,j}) \nn
&=& (\cosh K)^{N_b} \prod_{i,j} (1+\tanh K S_{i,j}S_{i,j+1})
(1+\tanh K S_{i,j}S_{i+1,j}),
\eea
where $N_b$ is the number of bonds.
In order to determine the partition function, we may expand this
expression in powers of $\tanh K S S'$ and sum over all spin configurations.
This summation yields zero unless all spins appear with even powers.
In this latter case the sum is $2^{N_s}$. This is the case, when the
interaction bonds form closed loops. That is at each lattice site
meet an even number of bonds as shown in the upper Figs.\ of Fig.\ 
\ref{Ising_square}.

The partition function can be expanded
\bea
Z(K) = 2^{N_s}(\cosh K)^{N_b} f(\tanh K), \label{Kr_xh} \\
f(a) =  \sum_l c_l a^l, \label{Kr_x}
\eea
where $K=\beta J$. The coefficients $c_l$ count the number of closed loops of
length $l$, $c_0=1$, $c_2=0$, $c_4=N_s$, $c_6=2N_s$, $c_8=N_s(N_s+9)/2$, etc.
and $c_l=0$ for odd $l$.
\smallskip

\noindent
{\bf Low temperature expansion} (\lte) We now consider the low
temperature expansion on the dual lattice. The dual lattice
is obtained by placing a spin $S^*(r^*)$ inside each of the squares
(in general polygons) of the original lattice. We multiply spins
$S^*$ in polygons with a common edge and sum over these products,
which in the case of the square lattice writes
\be
H^* = -J^* \sum_{i,j} (S^*_{i-\half,j-\half} S^*_{i-\half,j+\half}
+S^*_{i-\half,j-\half} S^*_{i+\half,j-\half}).
\ee
Assuming positive $J^*$ the states lowest in
energy are those where all $S^*$ are equal. Their energy is
\be
E^*_{min}=-N_b J^*,
\ee
where $N_b=2N^*_s$ is the number of bonds.

Excited states are found by turning some spins. Reversing one spin
costs an excitation energy $2lJ$, if the spin interacts with $l$ other
spins. Quite general the excitation energy is given by $2lJ$, if the
overturned spins are surrounded by Bloch walls of a total number of $l$
edges. In the case of the square lattice one obtains
\be
Z^*(K^*) = 2\ex{N_bK^*} f(\ex{-2K^*}) \label{Kr_xl}
\ee
with $f$ defined in (\ref{Kr_x}).
\smallskip

\noindent
{\bf Comparison}
Kramers and Wannier argued: If the partition function or equivalently
the free energy has a singularity at the critical point and no other
singularity, then it must be determined by
\be
\ex{-2K_c} = \tanh K_c,
\ee
which yields
\be
K_c = \frac 12 \ln(1+\sqrt 2) = 0.4407, \label{Kc}
\ee
which indeed turned out to be correct from Onsager's exact 
solution\cite{Onsager44}.
Thus there is a relation between the partition function and similarly the free
energy at high ($K<K_c$) and low ($K^*>K_c$) temperatures for
\be
\tanh K = \ex{-2K^*} \:\; \leftrightarrow \:\; \tanh K^* = \ex{-2K} \:\;
\rightarrow \:\; \sinh(2K) \sinh(2K^*) = 1.
\label{Kdual}
\ee

\begin{figure}[h]
\begin{center}
\includegraphics[scale=1.0]{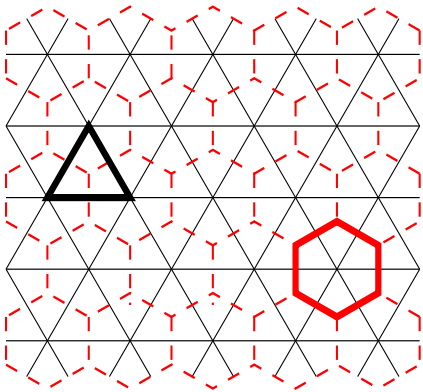}
\end{center}
\caption{Triangular and dual hexagonal lattice. The thick black triangle
indicates a product of 3 interactions on the triangular lattice
contributing to \hte\ and the Bloch wall for an overturned spin on the
hexagonal lattice. Similarly the thick red hexagon indicates a product
of 6 interactions on the hexagonal lattice contributiong to \hte\ and
the Bloch wall of an overturned spin on the triangular lattice.}
\label{Ising_hex2}
\end{figure}

The square lattice is called self-dual, since the \hte\ and the \lte\ are
performed on the same lattice. This is different for the triangular lattice,
where the \hte\ is performed on the triangular lattice and the \lte\ on the
honeycomb lattice. See Fig.\ \ref{Ising_hex2}. Then however the \hte\ of the
triangular lattice and the \lte\ of the honeycomb lattice are given by the
same sum $f(a)$,
\bea
Z^{\rm hte}_{\rm 3}(K) = 2^{N_{s3}} (\cosh K)^{N_{b}} f_3(\tanh K), &&
Z^{\rm lte}_{\rm 6}(K) = 2\ex{N_{b}K} f_3(\ex{-2K}), \\
Z^{\rm hte}_{\rm 6}(K) = 2^{N_{s6}} (\cosh K)^{N_{b}} f_6(\tanh K), &&
Z^{\rm lte}_{\rm 3}(K) = 2\ex{N_{b}K} f_6(\ex{-2K}),
\eea
where the number $N_b$ of bonds are equal in both lattices and $N_{s3}=N_b/3$
and $N_{s6}=2N_b/3$. The coefficients $c_l$ in $f_3$ and $f_6$  count the number
of closed loops on the triangular and the honeycomb lattice, resp.

As a consequence the partition functions $Z_3(K)$ and
$Z_6(K^*)$ are directly related for $K$ and $K^*$ given by (\ref{Kdual}).
One cannot directly read of the critical values $K_c$ for these lattices.
However, the Ising model on the honeycomb lattice can be related to that
on the triangular lattice by means of the star-triangle
transformation\cite{Wannier45}. To do this one eliminates every other spin
of the hexagonal lattice by summing $\sum_{S_0}\ex{KS_0(S_1+S_2+S_3)}$ one
obtains $C\ex{K'(S_1S_2+S_1S_3+S_2S_3)}$, which yields the Boltzmann factor of
the Ising model on the triangular lattice.

\section{Duality in 3 dimensions}
\label{threedim}

The basic question I asked myself, when I started my paper\cite{Wegner71}
on duality in generalized Ising
models, was: {\it Does there exist a dual model to the three-dimensional Ising
model?}
It turned out, that there is such a model, but of a different kind of
interaction. (Compare also \cite{BalDroItz75a})

In order to see this, I consider the low-temperature expansion of the 3d-Ising
model on a cubic lattice. I start out from the ordered state and then change
single spins.
These single spins are surrounded by closed Bloch walls.
The expansion of the partition
function is again of the form (\ref{Kr_x},\ref{Kr_xl}), but now with
$c_2=0$, $c_4=0$, $c_6=N_s$, $c_8=0$, $c_{10}=3N_s$, $c_{12}=N_s(N_s-7)/2$, etc.

\begin{figure}[h]
\begin{center}
\includegraphics[scale=0.5]{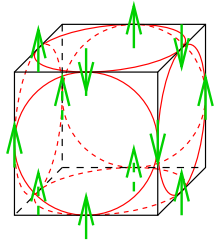}
\end{center}
\caption{Elementary cube with spins. The red (black) circles (ellipses)
indicate, which four spins are multiplied in the interaction}
\label{Ising_plaq2}
\end{figure}

The \hte\ of the dual model must be given by an interaction such
that only
closed surfaces yield a contribution. Thus locate a spin at each edge and
introduce
the interaction as a product of the spins surrounding an elementary square
called
plaquette. Thus the interaction of the dual model reads
\bea
H=-J\sum_{i,j,k}  &(& S_{i+\half,j,k+\half} S_{i+\half,j+\half,k} 
S_{i+\half,j,k-\half} S_{i+\half,j-\half,k} \nn
&+& S_{i+\half,j+\half,k}S_{i,j+\half,k+\half}S_{i-\half,j+\half,k}
S_{i,j+\half,k-\half} \nn
&+& S_{i+\half,j,k+\half}S_{i,j+\half,k+\half}S_{i-\half,j,k+\half}
S_{i,j-\half,k+\half}).
\eea
It is the sum over three differently oriented plaquettes. They are shown
in Fig.\ \ref{Ising_plaq2}.
\smallskip

\noindent
{\bf $S$-independent products of $R(b)$} From Fig.\ \ref{Ising_plaq2}
it is obvious that the product of the six $R(b)$s around the cube
does not depend on the spin configuration, since each spin appears twice
in the product.
\smallskip

\noindent
{\bf Gauge invariance} This model has a local gauge invariance. Turning
all spins around the corner of a cube does not change the energy of the
configuration. As an example in Fig.\ \ref{Ising_plaq2} the three spins
around the corner close to the center are reversed from the state,
in which all spins are aligned upwards.

\section{General Ising models and duality}
\label{gen}

\subsection{General Ising models}

We consider models with $N_s$ Ising spins on lattice sites $r$ described
by a Hamiltonian
\be
\beta H = -\sum_b K(b) R(b), \quad
R(b) = \prod_r S(r)^{\theta(b,r)}, \quad \theta(b,r)\in\Zm,
\label{Gen_Ham}
\ee
where $\Zm$ contains the two elements 0 and 1 of the ring modulo 2 with
\be
0+0=1+1=0, \quad 0+1=1+0=1, \quad 0\cdot 0= 0\cdot 1 = 1\cdot 0 = 0,
1\cdot 1=1.
\ee
We call the $b$ bonds; there number be $N_b$. The elements $\theta(b,r)$ of the
incidence 
matrix $\theta$ assumes the value 1, if $r$ belongs to the bond $b$, otherwise
0.
Thus $R(b)$ is the product of the Ising
spins $S(r)$ with $\theta(b,r)=1$. We denote the rank of the
matrix $\theta$ modulo 2 by $N_{\theta}$. Thus at least one
$N_{\theta}\times N_{\theta}$-subdeterminant equals one
modulo 2, whereas all $(N_{\theta}+1)\times(N_{\theta}+1)$-subdeterminants
equal zero modulo 2. If we write
\be
S(r)=(-)^{\sigma(r)}, \quad R(b)=(-)^{\rho(b)}, \quad
\sigma(r), \rho(b) \in \Zm,
\ee
then
\be
\rho(b) = \sum_r \theta(b,r)\sigma(r)
\ee
is the image of $\theta$.
If all $K(b)$ are positive, then one ground state is given by $S(r)=+1$.
In general there will be several ground states. They obey
$\rho(b)\zero$ for all $b$. These configurations $\sigma_0(r)$
constitute the kernel of $\theta$,
\be
\sum_r \theta(b,r)\sigma_0(r) \zero. \label{sigma}
\ee
There are $N_g=N_s-N_{\theta}$ linearly independent solutions $\{\sigma_0\}$.
which yield $2^{N_g}$ ground state configurations.

\subsection{Duality}

Besides the Ising model described by the Hamiltonian (\ref{Gen_Ham})
we consider a second Hamiltonian
\be
\beta^* H^* = -\sum_b K^*(b) R^*(b), \quad
R^*(b) = \prod_{r^*} S^*(r^*)^{\theta^*(b,r^*)}. \label{Gen_Hams}
\ee
with $N^*_s$ spins $S^*(r^*)$ on lattice sites $r^*$. The bonds $b$
are common to both Hamiltonians. Similarly we introduce the rank
$N^*_{\theta}$ and obtain the ground state degeneracy $2^{N^*_g}$
with $N^*_g=N^*_s-N^*_{\theta}$.

The two models are called dual to each other, if two conditions
are fulfilled:\\
(i) the closure condition
\be
\sum_b\theta(b,r)\theta^*(b,r^*) \zero \label{close}
\ee
for all pairs $r,r^*$, and\\
(ii) the completeness relation $N_m=0$, where
\be
N_m := N_b -N_{\theta}-N_{\theta^*}.
\ee
If these two conditions are fulfilled, and $K(b)$ and $K^*(b)$ are connected
by (\ref{Kdual}), then the partition functions of the two models are related
by
\be
Y\{K\} = Y^*\{K^*\}, \label{Dual}
\ee
where
\be
Y\{K\} = Z\{K\} 2^{-(N_s+N_g)/2} \prod_b (\cosh(2K(b))^{-1/2}
\label{ZY}
\ee
and similarly for $Y^*\{K^*\}$.
\smallskip

\noindent
{\bf Derivation of (\ref{Dual}, \ref{ZY})} The partition function $Z\{K\}$ can
be written in \hte
\bea
Z\{K\} &=& \sum_{\{S(r)\}} \ex{-\beta H}
=\sum_{\{S(r)\}} \prod_b \ex{K(b)R(b)} \nn
&=& \prod_b \cosh K(b) \sum_{\{\phi(b)\}} \prod_b \tanh K(b)^{\phi(b)}
\sum_{\{S(r)\}} \prod_b R(b)^{\phi(b)},
\eea
with $\phi(b)\in \Zm$ independent for all $b$.
Since
\be
\prod_b R(b)^{\phi(b)} = \prod_r S(r)^{\sum_b\theta(b,r)\phi(b)},
\ee
those $\phi(b)=\phi_0(b)$ contribute, which obey the set of homogeneous
equations
\be
\sum_b\theta(b,r)\phi_0(b)\zero \label{phi0}
\ee
for all $r$. Thus $\phi_0$ is the kernel of the transposed $\theta^{\rm t}$
of $\theta$.
Its dimension is $N_b-N_{\theta}$.
Thus there are in total $2^{N_b-N_{\theta}}$ solutions $\{\phi_0\}$.
They contribute with the factor $2^{N_s}$. Thus
\be
Z\{K\} = 2^{N_s} \prod_b \cosh K(b) \sum_{\{\phi_0(b)\}}
\prod_b \tanh K(b)^{\phi_0(b)}. \label{ZK}
\ee
The partition function $Z^*\{K^*\}$ reads in \lte
\bea
Z^*\{K^*\} &=& \sum_{\{S^*(r^*)\}} \ex{-\beta^* H^*}
= 2^{N^*_g} \sum_{{\rm closed} \{b\}} \prod_b \ex{K^*(b)R^*(b)} \nn
&=& 2^{N^*_g} \prod_b \ex{K^*(b)} \sum_{\{\rho^*\}}
\prod_b \ex{-2K^*(b)\rho^*(b)},
\eea
since $R^*(b)=1-2\rho^*(b)$. $\rho^*$ is the image of $\theta^*$,
\be
\rho^*(b) = \sum_{r^*} \theta^*(b,r^*) \sigma(r^*).
\ee
Its dimension is ${\rm dim}\,{\rm im}(\theta^*)=N_{\theta}^*$.
Due to the closure relation $\rho^*$ obeys the homogeneous equations
\be
\sum_b \theta(b,r)\rho^*(b)
= \sum_b\sum_{r^*}\theta(b,r)\theta^*(b,r^*)\sigma(r^*)\zero.
\ee
Thus $\rho^*$ belongs to the kernel of $\theta^{\rm t}$
with dimension ${\rm dim}\,{\rm ker}(\theta)=N_b-N_{\theta}$.
If both dimensions are equal,
${\rm dim}\,{\rm im}(\theta^*)={\rm dim}\,{\rm ker}(\theta)$
then the completeness relation is fulfilled, $N_m=0$, and
the partition function reads
\be
Z^*\{K^*\} = 2^{N_g^*}\prod_b \ex{K^*(b)} \sum_{\{\phi_0\}}
\prod_b \ex{-2K^*(b)\phi_0(b)} \label{Z*K*}
\ee
with the kernel $\phi_0$ of $\theta^{\rm t}$ as given in (\ref{phi0}).
The sums over $\{\phi_0\}$ are the same for $Z\{K\}$ and $Z^*\{K^*\}$
in (\ref{ZK}) and (\ref{Z*K*}).
Denoting
\be
f\{a\} := \sum_{\{\phi_0\}} \prod_b a(b)^{\phi_0(b)}, \quad
C:=\prod_b \frac{\cosh K(b)}{(\cosh(2K(b)))^{1/2}}
\ee
one obtains
\be
\prod_b \frac{\ex{K^*(b)}}{{(\cosh(2K^*(b)))^{1/2}}} = 2^{N_b} C
\ee
and
\bea
Y\{K\} &=& 2^{(N_s-N_g)/2} C f(\tanh K), \\
Y^*\{K^*\} &=& 2^{(N_b+N^*_g-N^*_s)/2} C f(\ex{-2K^*})
= 2^{(N_s-N_g+N_m)/2} C f(\ex{-2K^*}).
\eea
This yields the duality relation (\ref{Dual}) for $N_m=0$.

If $N_m>0$, then the summation in (\ref{Z*K*}) does not extend over the full
set $\{\phi_0\}$. Denote the sum (\ref{Z*K*}) by $f'\{a\}$ instead of $f\{a\}$,
then
\be
Y^*\{K^*\} = 2^{(N_s-N_g+N_m)/2} C f'(\ex{-2K^*}).
\ee
Since all terms in the sum $f$ are positive, one obtains $f'<f$ and thus
the inequality
\be
Y^*\{K^*\} < 2^{N_m/2} Y\{K\}.
\ee
We have obtained this relation from the \hte\ of $Z$ and the \lte\ of $Z^*$.
If instead we consider the \hte\ of $Z^*$ and the \lte\ of $Z$, then we
obtain a similar second inequality, in total
\be
2^{-N_m/2} Y\{K\} < Y^*\{K^*\} < 2^{N_m/2} Y\{K\}. \label{Yineq}
\ee
The difference in the free energy per lattice site vanishes in the thermodynamic
limit
due to the factors $N_m^{\pm 1/2}$ in (\ref{Yineq}),
if $N_m$ does not increase in the thermodynamic limit.
\smallskip

\noindent
{\bf Example: Two-dimensional Ising model}
The two-dimensional Ising model with $N_s$ spins on the square-lattice
yields $N^*_s=N_s$, $N_b=2N_s$, $N_g=N^*_g=1$, and thus $N_m=2$. The closed
loops which show up in the \hte, but not in the \lte are those, where
one loop runs around the torus in one or the other or both directions.
This corresponds to antiperiodic boundary conditions. Denoting the
partition function with
boundary conditions $S_{i,j}=s_xS_{i+N_1,j}=s_yS_{i,j+N_2}$ by $Z_{s_x,s_y}$
one obtains the exact relation
\be
Y\{K\} = \frac 12 (Y_{++}\{K^*\} + Y_{+-}\{K^*\} + Y_{-+}\{K^*\}
+ Y_{--}\{K^*\}).
\ee
The difference in the free energy per lattice site due to the factors
$N_m^{\pm 1/2}$ in (\ref{Yineq}) vanishes in the thermodynamic limit.

\section{Lattices and Ising-models}
\label{Mdn}

\subsection{Lattices and their dual lattices}

The models considered up to now are generalized to models in arbitrary 
dimensions $d$. I denote $k$-dimensional hypercells by $k$-cells.
I divide the $d$-dimensional hypervolume into $C_d$ $d$-cells $B^{(d)}$.
These are bounded by $(d-1)$-cells $B^{(d-1)}$.
Generally the $k$-cells $B^{(k)}$ are bounded by $(k-1)$-cells $B^{(k-1)}$.
Their number is denoted by $C_k$.
The 0-cells are simply the $C_0$ corners $B^{(0)}$ of the $d$-cells.

I associate lattice points $r^{(k)}$ to the
$k$-cells $B^{(k)}$. Their location will be specified more precisely below.

The dual lattice is obtained in the following way:
The points $r^{(d)}$ are the corners of the dual lattice.
Pairs of points $r^{(d)}$ are connected by 1-cells $B^{*(1)}$,
if the corresponding two cells are separated by a common $B^{(d-1)}$.
The 1-cells $B^{*(1)}$ crossing the cells $B^{(d-1)}$ around a given
cell $B^{(d-2)}$ is the boundary of a 2-cell $B^{*(2)}$.
Generally the $k$-cells $B^{*(k)}$ crossing the cells
$B^{(d-k)}$ around a cell $B^{(d-k-1)}$ is the boundary of a
$(k+1)$-cell $B^{*(k+1)}$. It is reasonable to define the
intersection of a cell $B^{*(k)}$ with its corresponding cell $B^{(d-k)}$
as the point $r^{*(k)}=r^{(d-k)}$. Thus the number of cells $B^{*(k)}$ equals
the number of points $r^{*(k)}$, $C^*_k=C_{d-k}$.

I define now the incidence matrix
\be
\theta(r^{(k+1)},r^{(k)}) = \left\{\begin{array}{cc}
1 & B^{(k)} \mbox{ on boundary } \partial B^{(k+1)}, \\
0 & B^{(k)} \mbox{ not on boundary of } \partial B^{(k+1)} \end{array} \right.
\ee
{\bf Closure relation:} An important property of the lattices is the closure
relation: Consider a pair 
$r^{(k+1)}$ and $r^{(k-1)}$. They lie in cells $B^{(k+1)}$ and $B^{(k-1)}$.
Then
\be
\sum_{r^{(k)}} \theta(r^{(k+1)},r^{(k)}) \theta(r^{(k)},r^{(k-1)}) \zero.
\label{closek}
\ee
Proof:
If $B^{(k-1)}$ is on the boundary of $B^{(k+1)}$, then two cells $B^{(k)}$
on the boundary of $B^{(k+1)}$ have $B^{(k-1)}$ as boundaries.
If $B^{(k-1)}$ is not at the boundary of $B^{(k+1)}$, then none of the $B^{(k)}$
on the boundary of $B^{(k+1)}$ has $B^{(k-1)}$ as boundary. This proofs
(\ref{closek}).

\subsection{Models on the lattice}

The model $M_{dn}$ have $C_{n-1}$ spins on lattice sites $r^{(n-1)}$
with an interaction defined by the bonds
\be
R(b)=\prod_{r^{(n-1)}} S(r^{(n-1)})^{\theta(r^{(n)}(b),r^{(n-1)})}.
\ee
The dual model $M^*_{d,d-n}$ has $C^*_{d-n+1}=C_{n-1}$ spins at
lattice sites $r^{(n+1)}$,
\be
R^*(b)=\prod_{r^{(n+1)}} S^*(r^{(n+1)})^{\theta(r^{(n+1)},r^{(n)}(b))}.
\ee
This defines together
with couplings $K$ and $K^*$ models (\ref{Gen_Ham}) and (\ref{Gen_Hams}).
Since there is a one-to-one corespondence between the bonds $b$ and the sites
$r^{(n)}$ I use interchangebly $b(r^{(n)})$ and $r^{(n)}(b)$.
\smallskip

\noindent
{\bf Gauge invariance} Changing all spins close to a point $r^{(n-2)}$,
\be
S(r^{(n-1)}) \rightarrow (-)^{\theta(r^{(n-1)},r^{(n-2)})} S(r^{(n-1)})
\ee
does not change the energy of the system, since any $R(b)$ is multiplied
by
\be
(-)^{\sum_{r^{(n-1)}} \theta(r^{(n-1)},r^{(n-2)})) 
\theta(r^{(n)}(b),r^{(n-1)})},
\ee
which due to the closure relation (\ref{closek}) yields one.
\smallskip

\noindent
{\bf Spin-independent products $R(b)$} The product over all $R(b)$ around
a given $r^{(n+1)}$, that is
\be
\prod_b R(b)^{\theta(r^{(n+1)},r^{(n)}(b))}
= \prod_{r^{(n-1)}} S(r^{(n-1)})^{\sum_{r^{(n)}} \theta(r^{(n)},r^{(n-1)})
\theta(r^{(n+1)},r^{(n)})} = 1.
\ee
does not depend on the spin configuration, since it yields one
due to the closure relation
(\ref{closek}). Of course also products of these products are
spin-independent.

\subsection{Euler characteristic and degeneracy}

{\bf Generalized Euler characteristic} The well-known Euler characteristic in
$d=2$ dimensions
\be
\chi = C_0-C_1+C_2,
\ee
where $C_0$ is the number of vertices (corners), $C_1$ number of edges,
and $C_2$ the number of faces, depends only on the topology of the surface.
For the plane one has $\chi=2$, if the outer face is also counted. For the
torus one has $\chi=0$. This characteristic can be generalized to arbitrary
dimension $d$,
\be
\chi = \sum_{m=0}^d (-)^m C_m. \label{chi}
\ee
Any lattice with the same boundaries (topology) can be created from any
other one by means of the following steps and their inverses:\\
{\bf Step:} An $m$-cell is divided into two such cells by
creating an $(m+1)$-cell between them. Then both
$C_m$ and $C_{m+1}$ increase by one and $\chi$ is conserved.

\begin{figure}
\begin{center}
\includegraphics[scale=0.8]{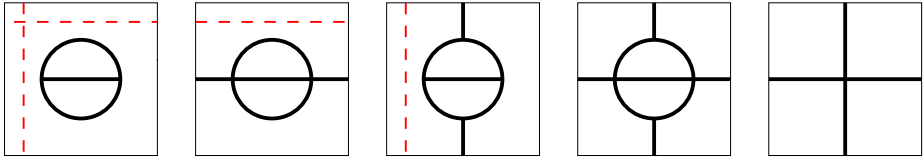}
\end{center}
\caption{Example of two-dimensional lattices in a periodicity square.
The first three examples do not
yield $\chi=0$, since walls indicated in red lines do not intersect
any edges. The last two examples yield $\chi=0$.}
\label{Ising_per}
\end{figure}

For periodic boundary conditions one obtains $\chi=0$, since we may cut
the lattice in one direction, double it and glue the two parts together.
Then all $C_m$ have doubled, and $\chi=2\chi$ and thus vanishes.
It is presumed that it is not possible to introduce an additional
'wall' $B^{(d-1)}$ in any periodic direction, which does not intersect
any of the original cells $B^{(d-1)}$. See Fig.\ \ref{Ising_per}.
\smallskip

\noindent
{\bf Degeneracy} 
We consider the change of $N_g$ resulting from the application of
a step as defined below (\ref{chi}): If $m>n$, then the Hamiltonian
is unchanged. For $m=n$ one bond and thus one interaction is duplicated
without change of degeneracy. For $m=n-1$ one spin is duplicated, but
for the ground state both must equal. For $m=n-2$ there is also one
additional spin. Taking this spin aligned upwards, one obtains again the
ground state. But changing the signs of all spins lying on bonds
adjacent to one $B^{(n-1)}$ at the boundary of the new bond, one
obtains another ground state. Then the system has twice the degeneracy
of the original system. The hamiltonian does not change for $m<n-2$.
Therefore we obtain
\be
N_g = t_g + \sum_{m=0}^{n-2} (-)^{n-m} C_m, \label{Ng1}
\ee
where $t_g$ depends only on the boundary condition.
Similarly one obtains
\be
N^*_g = t^*_g + \sum_{m=n+2}^d (-)^{m-n} C_m.
\ee
Thus
\bea
N_m &=& N_b -N_s + N_g -N^*_s + N^*_g 
= C_n - C_{n-1} - C_{n+1} + N_g + N^*_g \nn
&=& \sum_{m=0}^d (-)^{n-m} C_m +t_g + t^*_g
= (-)^n \chi +t_g+t^*_g.
\eea
We argue after (\ref{Ng2}) that
\be
t_g={d-1 \choose n-1}, \quad t^*_g = {d-1 \choose d-n-1}, \quad
N_m = {d \choose n}. \label{tg}
\ee
for periodic boundary conditions. Thus $N_m$ does not depend on the size of
the model.

\section{The models $M_{d,n}$ on hypercubic lattices}
\label{Mdncub}

We consider now the models $M_{d,n}$ on hypercubic lattices. The $k$-cells
are $k$-dimensional hypercubes with edges of unit length around $r^{(k)}$.
The lattice points $r^{(k)}$ have $k$ integer coordinates
and $(d-k)$ half integer coordinates, that is they are $\frac 12$ modulo 1.
They are defined by
\be
\begin{array}{cc}
r_i^{(k)}-\frac 12 < x_i < r_i^{(k)}+\frac 12 & r_i^{(k)}\in \mathbb{Z}, \\
x_i=r_i^{(k)} & r_i^{(k)} \in \mathbb{Z}+\frac 12.
\end{array}
\ee
The coordinates of the dual model are $r^{*(k)}=r^{(d-k)}$ and the
corresponding $k$-dimensional hypercubes are given by
\be
\begin{array}{cc}
r_i^{*(k)}-\frac 12 < x_i < r_i^{*(k)}+\frac 12 & r_i^{(k)}\in \mathbb{Z}+\frac
12, \\
x_i=r_i^{*(k)} & r_i^{*(k)} \in \mathbb{Z}.
\end{array}
\ee
We assume periodic boundary conditions, then
\be
C_k = C_{d-k} = C^*_k = C^*_{d-k} = {d \choose k} C_d.
\ee
The model $M_{dn}$ has spins on sites $r^{(n-1)}$.
The dual model $M^*_{d,d-n}$ has spins on sites $r^{(n+1)}=r^{*(d-n-1)}$.
Thus the model $M^*_{d,d-n}$ is the model $M_{d,d-n}$ shifted by 1/2 in all
coordinates.

Due to the above conditions $r^{(n+1)}$ and $r^{(n-1)}$ can only have bonds
in common, if they agree in $d-2$ coordinates and differ only in 2 coordinates.
Let these different coordinates be $(i,j)$ and $(i\pm\half,j\pm\half)$.
Then they have two bonds in common as claimed before: $(i,j\pm\half)$ and
$(i\pm\half,j)$ and fulfill the closure condition.
\smallskip

\noindent
{\bf Gauge invariance and degeneracy} If $n>1$, then one may reverse all spins
closest
to a given point $r^{(n-2)}$ without changing the energy of the system.
Thus these systems have a local gauge invariance. This leads to a high
degeneracy of the ground state. From (\ref{Ng1}) we obtain
\be
N_g = t_g + \sum_{m=0}^{n-2} (-)^{n-m} { d \choose m } C_d
= t_g + {d-1 \choose n-2} C_d \label{Ng2}
\ee
$t_g$ is determined by considering only one hypercube $C_d=1$ in the periodic
lattice. One obtains
$N_g=N_s$, since periodic boundary conditions
require that the spins in the products $R(b)$ are pairwise equal,
and we obtain $t_g$ as given in (\ref{tg}). Thus
\be
N_g = { d-1 \choose n-1 } + {d-1 \choose n-2} C_d.
\ee
Similarly one obtains
\be
N^*_g = {d-1 \choose d-n-1} + {d-1 \choose d-n-2} C_d,
\ee
and $t^*_g$ and $N_m$ as given in (\ref{tg}).
\smallskip

\nicht{
\noindent
{\bf Spin-independent products $\bf R(b)$} In the \hte we sum over
products of $R(b)$, which yield unity for arbitrary spin-configurations.
A large number of them are obtained by taking the product of
all $R(b(r^{(n)}))$, where the $r^{(n)}$ are neighbors of any
given $r^{(n+1)}$, since $S(r^{(n-1)})$ appears twice in the
product, if $r^{(n+1)}$ and $r^{(n-1)}$ are closest, otherwise
$S(r^{(n-1)})$ does not appear. Obviously also the products of
such 'elementary products' of $R(b)$ are spin-independent.
\smallskip
}

\noindent
{\bf Self-duality}
The model $M_{dn}$ on the hypercubic lattice is self-dual,
if $d=2n$. This is the case for $M_{2,1}$, which is the two-dimensional
Ising model on the square lattice. But also the four-dimensional model
$M_{4,2}$ with the plaquette interaction is self-dual. Both have the
phase transition at $K_c=0.4407$, (\ref{Kc}). The Ising model $M_{2,1}$
shows a continuous transition. Creutz, Jacobs, and Rebbi\cite{CrJaRe79a}
have investigated the model $M_{4,2}$ by Monte Carlo techniques. They
determined $\langle R(b)\rangle$ as a function of $K$. They found a
first order transition with hysteresis. By decreasing $K$ the system
showed superheating until $\approx 0.48$ and by increasing $K$ undercooling
until $\approx 0.40$. Starting from a mixed phase the phase transition
was located between $0.43$ and $0.45$.

Duality can be generalized to Abelian groups $Z(N)$. Let
$S(r)=\ex{2\pi\ie p/N}$ with $p=0,...N-1$ and the energy assigned to the
product of two spins in states $p$ and $p'$ by $E_{p-p'}$, then
the weights $\omega_{p-p'}=\ex{-\beta E_{p-p'}}$ and their dual are related
by the Fourier transform\cite{Wegner73}
\be
\omega^*_p = N^{-1/2} \sum_{p'} \ex{2\pi\ie pp'/N} \omega_{p'}.
\ee
This can be generalized to the models $M_{dn}$. The models $M_{42}$ are
self-dual for $Z_N$ with $N=3,4$ and the critical $K_c$s are
determined\cite{Kort78,Yoneya78}.
Monte-Carlo calculations\cite{CrJaRe79b} confirm these transition temperatures
for $N=3,4$.
Corresponding calculations yield two phase transitions for $N\ge 5$.
For more general aspects of duality in Abelian groups see section 6.1.4
Duality in\cite{ItzDro}.

\section{Correlations}
\label{correl}

Non-vanishing correlations are only obtained for gauge-invariant products.
These are products of $R(b)$. In particular we consider the product of
spins on the boundary of an $n$-dimensional hypercube of $M_{dn}$.
The \hte\ yields
\bea
\langle \prod_r S(r) \rangle &=& (\tanh K + 2(d-n)(\tanh K)^{1+2n}+...)^{\vau},
\quad n>1, \\
&=& \frac 12[\tanh K +(2(d-1))^{1/2} (\tanh K)^2 +...]^{\vau} \nn
&+& \frac 12[\tanh K +(2(d-1))^{1/2} (\tanh K)^2 +...]^{\vau}, \quad  n=1.
\eea
where $\vau$ is the volume of the hypercube. For $n=1$ this is the distance
between the two spins; for $n=2$ it is the area spanned by the spins.
The \lte\ yields 
\bea
\langle \prod_r S(r) \rangle = (1-\ex{4(d-n+1)K} + ... )^f, \quad n<d, \nn
\langle \prod_r S(r) \rangle = (1-2\ex{-2K} +... )^{\vau}, \quad n=d,
\eea
where $f$ is the hyperarea of the boundary of the hypercube
(for $n=1$ it is the number $f=2$ of ends of the line; for $n=2$, $f$ is
the perimeter of the square). Thus the behavior of the correlation functions
of large hypercubes differs in the high and low temperature phases, and we
expect
\be
\langle \prod_r S(r) \rangle \propto \left\{ \begin{array}{cc}
\ex{-\vau/\vau_0(T)} & T>T_c, \, n<d \\
\ex{-f/f_0(T)} & T<T_c, \, n<d \end{array} \right. \label{corr}
\ee
We attribute the qualitatively different asymptotic behavior in both temperature
regions to different states of the system above and below a critical temperature
$T_c$.
\smallskip

\noindent
{\bf The model $\bf M_{dd}$} The only restriction on the $R(b)$ is that the
product
of all of them equals one. Consequently the partition function reads
\be
Z(K) = 2^{N_s} [(\cosh K)^{N_b} + (\sinh K)^{N_b}]
\ee
The expectation value of a product of $\vau$ factors $R$ yields
\be
\langle \prod_b R(b) \rangle
= \frac{(\tanh K)^{\vau} + (\tanh K)^{N_b-\vau}}{1+(\tanh K)^{N_b}}.
\ee
The models $M_{dd}$  do not show a phase transition. Among these models is
$M_{11}$, a closed linear chain of Ising spins.

\subsection{Dislocations}

We consider systems with magnetic dislocations. Let the operator $M(b)$ change
the sign of $K(b)$. We introduce $\phi^*(b)=1$ for bonds with changed signs, and
$\phi^*(b)=0$ for bonds with unchanged coupling. Then the expectation value
of the product of $M(b)$s is
\bea
&& \langle \prod_b M(b)^{\phi^*(b)} \rangle
= \langle \prod_b \ex{-2\phi^*(b)K(b)R(b)} \rangle \nn
&& = \frac{Z\{(-)^{\phi^*}K\}}{Z\{K\}}
= \frac{Y\{(-)^{\phi^*}K\}}{Y\{K\}}. \label{dis1}
\eea
From (\ref{Kdual}) we obtain $\tanh((-)^{\phi^*}K) = \ex{-2K^*-\ie\pi\phi^*}$
and thus
\bea
&& \langle \prod_b M(b)^{\phi^*(b)} \rangle
= \frac{Y\{K^*+\ie\pi\phi^*/2\}}{Y\{K^*\}} \nn
&=& \ie^{-\sum_b\phi^*(b)}\langle \prod_b \ex{\ie\pi\phi^*(b)R^*(b)/2}
\rangle\{K^*\} = \langle \prod_b R^*(b)^{\phi^*(b)}\rangle\{K^*\}. \label{dis2}
\eea
\smallskip

\begin{figure}[h]
\begin{center}
\includegraphics[scale=1.0]{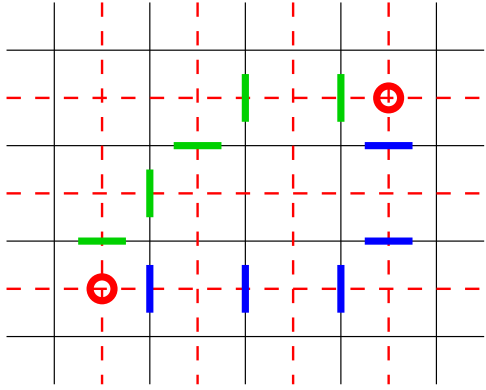}
\end{center}
\caption{Correlation between the two spins at the dots}
\label{Ising_dis}
\end{figure}

\noindent
{\bf Interpretation} Kadanoff and Ceva\cite{KadCeva71} introduced this
concept for the two-dimensional Ising model $M_{2,1}$. Take a sequence
of bonds $b$ indicated by either the blue (black) or the green (grey)
bars between the two spins at the sites indicated in Fig.\
\ref{Ising_dis} by two red (black) circles, then
$\langle \prod_b R^*(b) \rangle$
is the product of these two Ising spins at $K^*$.
It equals the ratio of the partition
functions with the changed bonds and the unchanged bonds,
thus the exponential of the difference $\Delta F$ of the free energy
without and with the changed bonds at $K$,
\be
\langle S(r^*) S(r^{*'}) \rangle(K^*) = \ex{-\Delta F(K)}.
\ee
If $K$ is in the paramagnetic region, then the disturbance of the bonds
yields a contribution to $\Delta F$ only close to the points, where this
line of bonds ends.
Thus for large separation of the two spins it approaches a finite
value, which corresponds to the square of the magnetization at $K^*$.
On the other hand if $K$ is in the ferromagnetic region, then the
disturbance will change the free energy proportional to the distance
between the two spins $S(r^*)$ and $S(r^{*'})$, which yields an
exponential decay of the correlation function.

Let us now consider $M_{3,1}$ and $M_{3,2}$. Change the sign of
the interaction $\sum_{ij} S_{ijk} S_{ijk+1}$ over a whole region
(area) in the plane spanned by $ij$. Analogous to the two-dimensional Ising
model, the change $\Delta F(K)$ will be proportional to the perimeter $f$
for paramagnetic $K$ and proportional to the area $\vau$ for ferromagnetic $K$.
The product $\prod_b R^*(b)$ is now the product of the Ising spins
along the perimeter of the dislocations. Consequently the expectation
value decays proportional to $\ex{-f/f_0(T^*)}$ at low temperatures $T^*$
and proportional to $\ex{-\vau/\vau_0(T^*)}$ at high temperatures $T^*$ in
accordance with (\ref{corr}).
\smallskip

\noindent
{\bf Local order parameter} If all states are taken into account. the
correlations different from zero are only obtained from products of $R$,
For $n=1$ the product of two spins $S(0)S(r)$ can be written as product
of $R$s. For $n>1$ products of spins $\prod_k S(a_k) \prod_l S(r+a_l)$ with
with $a_k$ and $a_l$ restricted to some finite region $|a_k|<c$, $|a_l|<c$
yield only non-vanishing correlations for distances $r>2c$, if both $\prod_k
S(a_k)$
and $\prod_l S(r+a_l)$ are separately gauge invariant, that is, they are
expressed as
finite products of $R$. However, with (\ref{dis1}, \ref{dis2}) expectations of
products
of $R$ in one phase can be expressed by correlations in the other phase
\be
\langle \prod_{{\rm some}\,b} R(b) \rangle\{K\}
=  \langle \prod_{{\rm same}\,b} (\cosh(2K^*(b))-R^*(b)\sinh(2K^*(b)))
\rangle
\ee
Thus since there is no long range order in the high temperature phase, there can
be none in the low temperature phase,
\be
\lim_{r\rightarrow\infty} (\langle\prod_k S(a_k) \prod_l S(r+a_l)\rangle
-\langle \prod_k S(a_k)\rangle \langle \prod_l S(a_l)\rangle ) = 0.
\ee
Thus there is no local order parameter for models $M_{dn}$ with $n>1$.
This argument does not apply for $n=1$, since in this case the number of $R$s
in the product increases with $|r|$.

\section{Lattice gauge theories}
\label{latt}

We have seen that models $M_{dn}$ with $n>1$ show local gauge invariance.
Such models are related to quantum chromodynamics. The basic idea first
formulated by Wilson\cite{Wilson74}
is to start from the lattice, we introduced as $M_{42}$.
(For a retrospect by Wilson see
\cite{Wilson04}. Many reprints on this subject are compiled in Rebbi's
book\cite{Rebbi83}).
The degrees of freedom are now denoted by $U$ in place of $S$.
These $U$ are elements of a group. It may be a finite or a continuous group,
it may be an Abelian or non-Abelian group. In the case of QCD one considers
the 'colour'-group SU(3). Let us denote the $U$ placed on the link between 
lattice sites $i$ and $j$ by $U_{ij}$, where one requires
$U_{ji}=U_{ij}^{-1}$. The action is a sum of terms
\be
g^{-2} \sum_{\rm plaquettes} (1-\frac 1N \Re{\rm tr}(U_{ij}U_{jk}U_{kl}U_{li})),
\ee
where $N$ is the dimension of $U$.
In addition one introduces quarks (fermions) with interaction
\be
g^{\prime-2} \sum_{\rm links} \psi_i\dg U_{ij} \psi_j.
\ee
These interaction terms are invariant under local gauge transformations
\be
\psi_j \rightarrow G_j \psi_j, \quad
\psi\dg_j \rightarrow \psi\dg_j G\dg_j, \quad
U_{ij} \rightarrow G_i U_{ij} G\dg_j.
\ee
The couplings depend on temperature and pressure of the hadron system.
At low temperature and pressure the correlations fall of with an
area law. Since the action is an integral over time, this behaviour
corresponds to an increase of the effective potential between quarks
proportional to the distance between them.
The gradient of the potential is called string tension and given by
$1/\vau_0(T)$ in (\ref{corr}). This potential binds three quarks,
which constitute a hadron. Or one quark and one antiquark
are bound and constitute a meson. Generally the difference between
the number of quarks and antiquarks has to be a multiple of three.
At high temperature and high pressure the system forms a quark-gluon
plasma. This corresponds to the phase in which the correlation increases
proportional to the perimeter of the loop. Then the effective potential
between the quarks stays finite at large distances and the quarks are
rather free to move in this plasma.

\section{Electromagnetic field}
\label{electro}

The electromagnetic field in QED and its coupling to charged particles can
be described similarly with the group $U(1)$,
\be
U_{ij} = \ex{\ie\int_j^i A_{\mu} \de x^{\mu}}
\ee
Then
\bea
&& {\rm tr}( U_{r,r+a^{\mu}e_{\mu}}
U_{r+a^{\mu}e_{\mu},r+a^{\mu}e_{\mu}+a^{\nu}e_{\nu}}
U_{r+a^{\mu}e_{\mu}+a^{\nu}e_{\nu},r+a^{\nu}e_{\nu}}
U_{r+a^{\nu}e_{\nu}} ) \nn
&& \approx \ex{\ie a^{\mu}a^{\nu} 
F_{\mu\nu}(r+(a^{\mu}e_{\mu}+a^{\nu}e_{\nu})/2)}
\eea
with the electromagnetic field tensor
\be
F_{\mu\nu} = \partial_{\mu}A_{\nu} - \partial_{\nu}A_{\mu}.
\ee
Since only the real part of ${\rm tr}(\prod U)$ contributes, one
obtains in leading order the well-known action of the
electromagnetic field proportional to $F_{\mu\nu}F^{\mu\nu}$.
If one performs the continuum limit ($a\rightarrow 0$) then only these terms
survive.

The discretized Maxwell equations can be solved on such a
lattice\cite{Weiland77}. One places the components $A_{\mu}$
on sites $r^{(1)}$, the six electromagnetic field components
$F_{\mu\nu}$ on sites $r^{(2)}$, the components of the charge
and current densities on sites $r^{(1)}$. Lorenz gauge
and charge conservation can be put on sites $r^{(0)}$.

\end{document}